\documentclass[12pt,preprint]{aastex}
\newcommand{\al} {$^{26}$Al}
\newcommand{\uthree}  {$\times 10^{-3}$ ph~cm$^{-2}$~s$^{-1}$}
\newcommand{\ufour}  {$\times 10^{-4}$ ph~cm$^{-2}$~s$^{-1}$}
\newcommand{\ctwo}  {$\times 10^{-2}$ counts~s$^{-1}$}
\newcommand{\e}  {1809~keV}
\begin{document}

\title{The \it Reuven Ramaty High Energy Solar Spectroscopic Imager \rm observation of the 1809~keV Line from Galactic $^{26}$Al}

\author{D. M. Smith\altaffilmark{1}}

\altaffiltext{1}{Space Sciences Laboratory, University of California Berkeley, 
Berkeley, CA 94720}

\begin{abstract}

Observations of the central radian of the Galaxy by the \it Reuven
Ramaty High Energy Solar Spectroscopy Imager (RHESSI) \rm have yielded
a high-resolution measurement of the \e\ line from \al, detected at
11$\sigma$ significance in nine months of data.  The \it RHESSI \rm
result for the width of the cosmic line is $(2.03 ^{+0.78}_{-1.21})$~keV
FWHM.  The best fit line width of 5.4~keV FWHM reported by
\citet{Na96} using the Gamma-Ray Imaging Spectrometer (GRIS) balloon
instrument is rejected with high confidence.

\end{abstract}


\section{Introduction}

The \e\ line of \al\ was the first astronomical gamma-ray line from
radioactive decay discovered in the Galaxy \citep{Ma84}, confirming
a prediction made by \citet{Ra77} and \citet{Ar77}. It is the
brightest Galactic line after the line from positron annihilation at
511 keV.  This first result was from \it HEAO~3\rm, which used
high-resolution germanium detectors.  
\citet{Ma84} found that the line
width was consistent with their instrumental resolution of about
3~keV, but the statistical significance of the result was low
(4.8~$\sigma$).

This isotope is thought to be produced by proton capture on magnesium, and
therefore can occur in any site of nucleosynthesis where these two
components are abundant \citep{Pr96}.  Environments suggested for its
creation and release into the interstellar medium
include type-II supernovae, novae,
and the winds of Wolf-Rayet and Asymptotic Giant Branch
stars.  Because its half-life is
around 10$^{6}$ yr, much shorter than scales of Galactic evolution,
its distribution on the sky reflects the current Galactic distribution
of the relevant parent events.  Maps of the Galactic \e\ emission
\citep[][and references therein]{Kn99}
were made with the COMPTEL instrument on the \it Compton Gamma-Ray
Observatory \rm and have given
us the first detailed look at where the emission is concentrated.  The
maps correlate better with early than late stellar populations,
suggesting that supernovae and/or Wolf-Rayet stars are likely to be
the primary contributors.

Many other observations of this line, with high- and low-resolution
balloon instruments and with low-resolution satellite instruments,
have occurred since \it HEAO~3 \rm (see \citet{Pr96} for a
comprehensive review).  Only one high-resolution measurement has had a
statistical significance comparable to the \it HEAO~3 \rm data and thus
been able to further advance our understanding of the shape of the
line.  This was performed by the Gamma-Ray Imaging Spectrometer (GRIS)
balloon \citep{Na96}.  The authors found the line to be significantly
broadened, with an intrinsic width of $(5.4 ^{+1.4}_{-1.3})$~keV derived
from a measured width of $(6.4 ^{+1.2}_{-1.1})$~keV by subtracting their
instrumental width of $(3.4 \pm 0.1)$~keV in quadrature.  The
significance of their overall detection was $6.8\sigma$, slightly
higher than that of \it HEAO~3\rm.

This Doppler broadening corresponds to isotropic velocities of 
540~km~s$^{-1}$ or a temperature of $\sim 4.5 \times 10^{8}$K
\citep{St99}, and there is no model for how either could be maintained
by a gas in the interstellar medium for anything approaching the
lifetime of the isotope.  This result has stimulated interesting
theoretical work centered on concentrating the \al\ in grains, which
can maintain their birth velocities much longer than gaseous material
and even be re-accelerated in supernova shocks
\citep{Ch97,El97,Li98,St99}.

\section{The instrument and analysis technique}

The \it Reuven Ramaty High Energy Solar Spectroscopic Imager (RHESSI)
\rm is a NASA Small Explorer satellite in a nearly circular $\sim$600~km
orbit (96--minute period) with inclination $\sim$38$^{\rm{o}}$.  Its
primary mission is to make high-resolution images and spectra of solar
flares in the range from 3~keV to 17~MeV \citep{Li02, Li03}.  Its
detector array is a set of nine high-purity coaxial germanium detectors
cooled to liquid nitrogen temperature by a Stirling-cycle refrigerator
\citep{Sm02}.  Each of \it RHESSI\rm's detectors is segmented into a
thin front segment facing the Sun (to stop hard x-rays) and a much
thicker rear segment meant to record solar gamma-rays that penetrate
the front segment.  Because the front segments have little effective
area at \e\ for any incident angle, I did not use them for this
analysis.  One of the nine detectors operates in an unsegmented mode
with poor energy resolution, and is excluded as well.  The spacecraft
rotates at close to 15~rpm about the axis pointing at the Sun.

Because the array is unshielded and the spacecraft is very light, the
effective area of the array to highly penetrating \e\ photons is
nearly independent of the direction of incidence, either in azimuth or
zenith angle with respect to the spacecraft axis.
This has been verified with Monte Carlo
simulations using the GEANT3 package with a highly detailed model of
the spacecraft, which show a maximum deviation of $\pm$10\% at
any angle from the mean effective area averaged over all angles.
Thus the annual rotation of the spacecraft with
respect to the stars cannot be used to modulate the Galactic signal as
was done with data from the \it Solar Maximum 
Mission \rm Gamma-Ray Spectrometer \citep{Ha90}.
As an additional complication, there is a background line in the
instrument at \e\ due to cosmic-ray interactions with aluminum in the
spacecraft.  This must be understood and subtracted before the
Galactic line can be studied.

The lack of directionality and strong background line can be overcome
by using the Earth as an occulter.  To begin the analysis, I divided
nine months of RHESSI data (1 March 2002 to 2 December 2002) into
one-minute (and therefore spin-averaged) intervals.  I defined the
``inner Galaxy'' as a box running from $\pm 30^{\rm{o}}$ in Galactic
longitude and $\pm 5^{\rm{o}}$ in Galactic latitude.  ``Source''
intervals were times when this entire box was unocculted by the Earth,
and ``background'' intervals were defined as times when it was
entirely occulted.  The rest of the data were discarded.  Because the
Earth subtends less than half the sky, the set of source pointings is
larger. The total amount of time accumulated in the source spectra
is 75.0 dy, and the total accumulation of background data
is 33.7 dy.  Data contaminated by the precipitation of magnetospheric
electrons were identified with the onboard particle detector and
were removed from consideration, as were data taken during the
X4.8 solar flare of 23 July, the only flare observed so far with
emission at energies approaching 1809~keV. 

The background subtraction algorithm is adapted from one
used to derive the most precise value to date for the flux
of positron-annihilation radiation from the Galactic Center region:
$(1.49 \pm 0.02)$\uthree\ \citep{Sm97}.  In that case the instrument
was the Burst and Transient Source Experiment (BATSE) on the \it
Compton Gamma-Ray Observatory (CGRO)\rm.  The principle is to sort the
background spectra into a two-dimensional library based on the two
most important parameters that control the background.  For each
source spectrum, a background spectrum is generated by selecting among
the library spectra using the values of the controlling parameters for
that source spectrum.  The computer code tracks the repeated
contributions of each original raw background spectrum in order to
propagate statistical errors correctly.

All background lines with half-lives less than a few hours are fairly
well controlled by choosing two particular parameters on which to sort
the library: the longitude of the ascending node (LAN) of the current
orbit, and the orbital phase since the ascending node (PAN).  These
parameters define a particular spot on the Earth, so that the
cosmic-ray flux (which is controlled primarily by geomagnetic
latitude) is reproduced.  The lines with half-lives of seconds or less
are proportional to the instantaneous cosmic-ray flux: this includes
the \e\ background line due to cosmic-ray production of $^{26*}$Mg in
aluminum, which decays with a 0.49~ps half-life \citep{Wh89}.  Unlike
simple geographic or geomagnetic coordinates, however, LAN and PAN
also contain a memory of the recent history of the spacecraft
trajectory -- most usefully, how recently and how deeply the
spacecraft passed through the SAA.  Thus matching LAN and PAN also
produces a good match for the SAA-induced background lines, which have
half-lives from minutes to hours.  We see such a line around 1811~keV
that blends with the \e\ line.  The only candidate in the gamma-ray
tables of \citet{Ch99} is the 1810.77~keV line of $^{56}$Mn (half-life
2.6~hr), presumably created by interactions of SAA protons with iron
near the detectors.  This line was suggested as a contributor to the
background of the germanium spectrometer on the \it Wind \rm
spacecraft \citep{We02}, and the published \e\ background line from
\it HEAO~3 \rm \citep{Ma84} shows an excess in the blue wing that is
consistent with the $^{56}$Mn line appearing at a low level.  There is
no very long-lived background line (such as the 1275~keV line from
$^{22}$Na) near \e.  Such a line would require a background-selection
system which is cognizant of the total time since the start of the
mission, rather than relying entirely on LAN and PAN.

\section{Results}

\begin{figure}
\epsscale{1.}
\plottwo{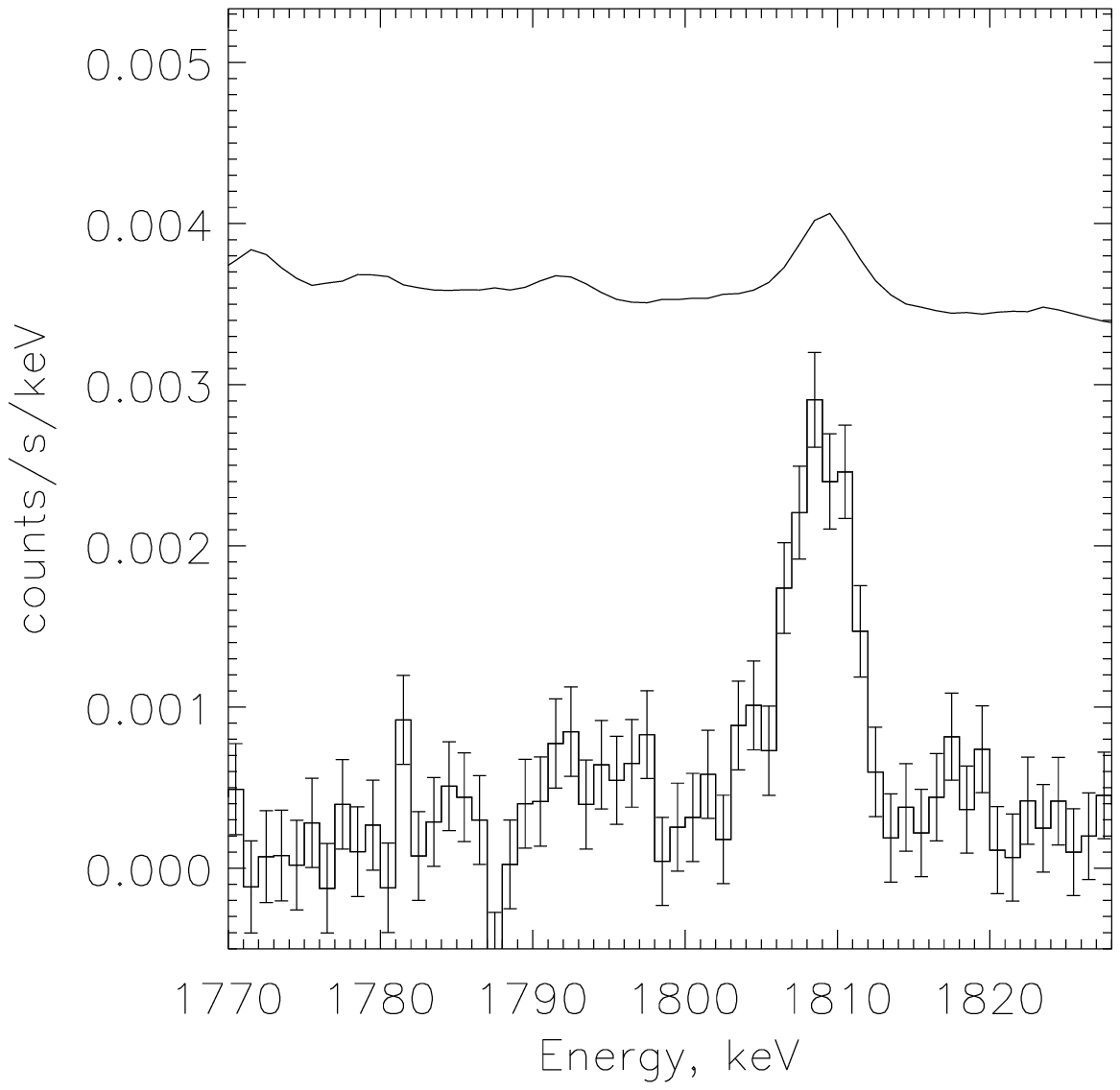}{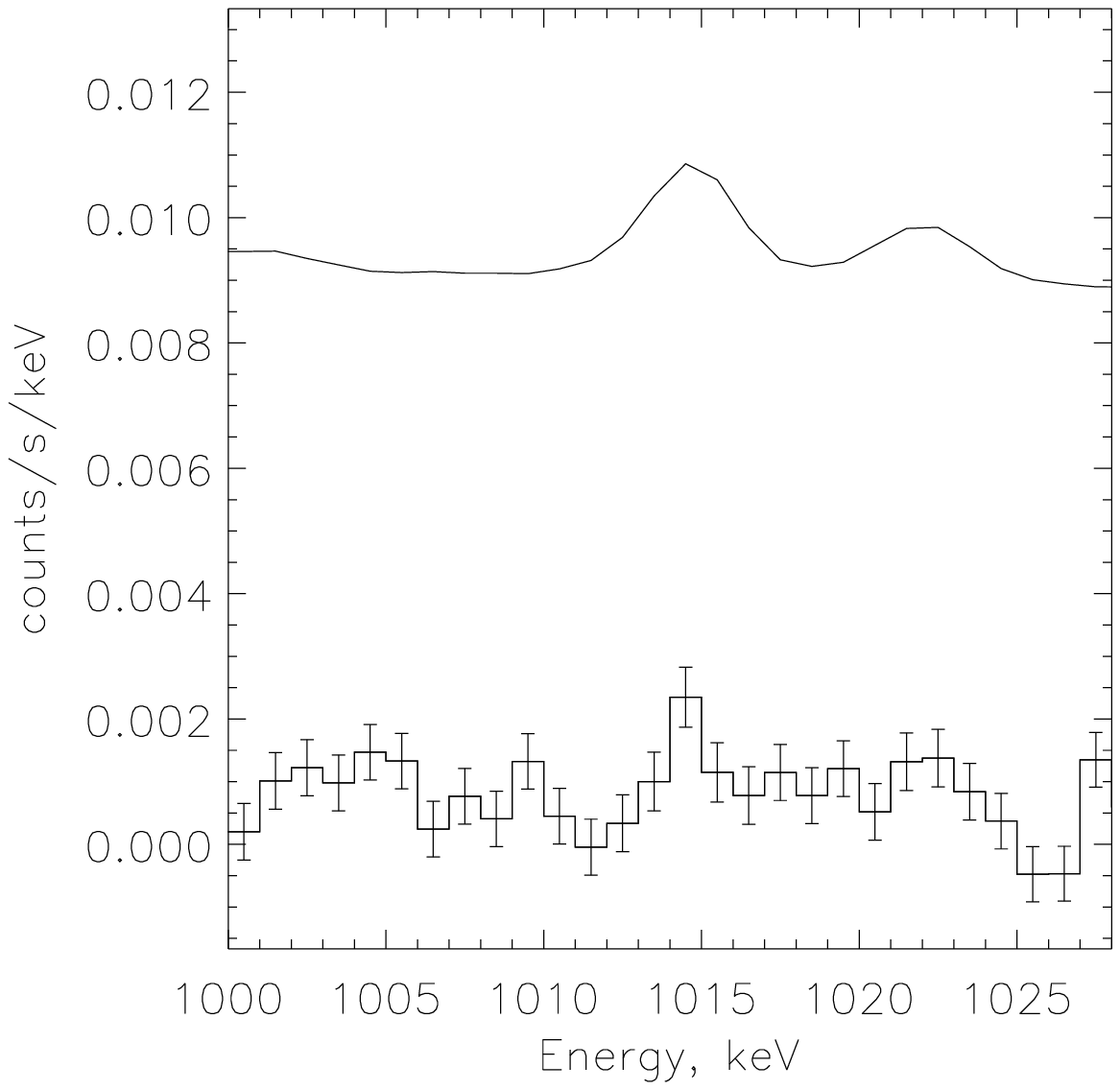}
\caption{ Inner-Galaxy spectra in the vicinity of
the 1809~keV line (left) and the background line at 1014~keV (right).
In each plot, the lower curve is the 
\it RHESSI \rm background-subtracted count spectrum and the upper
curve is 3\% of the average
background spectrum during this period. The bins shown are 1~keV wide
and the instrumental resolution at 1809~keV is $(4.10 \pm 0.07)$~keV
(see text).}
\end{figure}

Figure~1 (left) shows the RHESSI background-subtracted spectrum of the
Galactic \al\ line.  The smooth curve shown is 3\% of the average
background for comparison.  Fitting the spectrum from 1790--1825~keV
to a Gaussian plus a linear background, I find a center energy of
$(1808.87 \pm 0.18)$~keV, 1.2$\sigma$ from the expected rest energy of
the line at 1808.65~keV \citep{Ch99}.
The area of the line is $(1.17 \pm 0.11)$\ctwo\
(11$\sigma$), which is 12.9\% of the background line (averaged
over the entire set of background spectra interpolated from
the background library).
Its FWHM, which includes instrumental
broadening, is $(4.58 \pm 0.44)$~keV.  The fit is good
($\chi^2$ = 26.7 with 30 degrees of freedom).

By fitting seven narrow background lines ranging from 186~keV to 2243~keV,
I interpolate the intrinsic instrumental resolution to its value
at \e\ and find $(4.10 \pm 0.07)$~keV.  I avoid the instrumental
background line at \e\ itself because of the blend with the line
at 1811~keV, which makes it much broader than the interpolation
($(4.86 \pm 0.08)$~keV FWHM) and raises its center energy to
$(1809.34 \pm 0.03)$~keV.  Subtracting the instrumental resolution
in quadrature from the Galactic line gives an intrinsic width
of $(2.03 ^{+0.78}_{-1.21})$~keV, corresponding to velocities
of $\sim$200~km/s or a temperature of $1.7 \times 10^{8}$K.

To demonstrate the quality of the background subtraction, I show in
Figure~1 (right) the region around a background line at 1014~keV.  This line,
like the background line at \e, is due to the prompt decay of a
short-lived isomer created by cosmic-ray interactions in aluminum.
Thus, if it subtracts well, one should be able to assume that the \e\
background line does also.  The figure demonstrates subtraction better than
the 3\% shown for comparison.  There is a hint of a residual 1014~keV
line which, when fit with a Gaussian, gives a flux of $(1.2 \pm
0.5)$\% of the background line.  The averaged cosmic-ray count rate in
the detectors over all the periods used for source pointings is 0.46\%
higher than the same average over the background periods.  This is
consistent with the slight undersubtraction hinted at in 
Figure~1 (right).  I
find that subtracting out an extra 0.46\% or 1.2\% of the \e\
background line has no significant effect on the Galactic \e\ line
width, whether I use the natural form of the background spectrum
(which includes the 1811~keV line) or whether I substitute an
artificial single Gaussian line at exactly 1808.65~keV with the instrumental
resolution.  Since the Galactic line flux is 12.9\% of the average
background, these extra subtractions would reduce its value by 4\% and
9\%, respectively.

To further demonstrate that the cosmic-ray-induced background line is
well-subtracted, I divided the data set into two roughly equal parts
by the value of the cosmic-ray count rate during each one-minute
spectrum.  The rate averaged 158 counts s$^{-1}$ for the lower data
set and 237 counts s$^{-1}$ for the higher data set, for a ratio of
1.50.  The count rate in the total, unsubtracted \e\ line had a
similar ratio of 1.42, with 8.00\ctwo\ and 11.36\ctwo\ in the two sets,
respectively.  Once the backgrounds were subtracted, however, the
values for the residual (Galactic) line were $(1.13 \pm 0.15)$\ctwo\
and $(1.23 \pm 0.15)$\ctwo, respectively, a ratio of only 1.09 and
well within statistical agreement.  The derived Galactic signal thus
appears to be independent of the intensity of the background line.

The conversion of the count rate in the Galactic line to a total
Galactic flux is strongly dependent on the assumed Galactic
distribution.  The effective area of the eight rear segments used,
calculated at \e\ using the distribution of angles of incidence over the
observation period assuming the source is concentrated at the Galactic
center, is 20.5~cm$^{2}$.  The incident flux is then $(5.71 \pm
0.54)$\ufour\ for the artificial case of a point source.  Since the
effective area of the instrument is nearly independent of angle, the
model distribution is important in determining the flux only because
of occultation of parts of the distribution by the Earth.  As noted
above, I required the inner 60$^{\rm{o}}$ of Galactic longitude to be
visible in each pointing, but on average a considerable additional
amount of the plane is also included.  A certain amount of this
high-longitude plane emission also appears in the background
intervals, however, and is subtracted off.

\section{Discussion}

Adding the GRIS best-fit value of 5.4~keV for the Galactic line width
in quadrature with \it RHESSI\rm's instrumental resolution would give
a 6.78~keV width.  Fixing the width at that value for the fit from
1790--1825~keV causes $\chi^{2}$ to increase by 17.3 to 44.0 (now with
31 degrees of freedom).  The probability of \it RHESSI\rm's result
being consistent with the GRIS best fit is then 4$\times 10^{-5}$ for
one parameter of interest.

The \it RHESSI \rm value for the Galactic line's width is only
marginally consistent with zero.  It is expected, however, that the
interstellar \al\ will share in Galactic rotation and display
appropriate Doppler shifts \citep{Sk91,Ge96}.  \citet{Ge96} combined a
three-dimensional model \al\ distribution derived from COMPTEL data
with a Galactic rotation model to produce a map of the radial velocity
versus Galactic longitude (integrated over Galactic latitude and line
of sight).  This map has a Galactic bulge component with velocities up
to $\sim$75 km/s and spiral arm components with velocities up to
$\sim$150 km/s.  Although an integration of this map over the inner
Galaxy has not been published, from the slices shown at different
longitudes it appears that a integrated FWHM of not more than 150 km/s
or about 0.9~keV would be obtained.  This is within 1$\sigma$
of our result.  Additional broadening due to more
local motions of the \al-bearing gas or dust is possible, and of
course the most recently-created portion of the \al\ can be traveling
near its birth velocity in supernovae, etc., even if this portion is
small.
 
\it RHESSI\rm's flux value of $(5.71 \pm 0.54)$\ufour\ is comparable
to previous measurements.  A value of $\sim 4$\ufour\ rad$^{-1}$ was
considered consistent with the existing ensemble of data sets at the
time by \citet{Di97}.  For a uniform distribution in the plane, this
would imply about 1.4~rad for \it RHESSI\rm's effective field of view
(including the effect of subtracted flux in the background pointings).
Realistic distributions fall off with longitude, however, and would
require an even larger effective field of view.  \citet{Na98},
however, using the GRIS data, found higher fluxes per radian, with the
value depending on the model used and ranging up to $(5.48 \pm
0.78)$\ufour\ rad$^{-1}$ assuming the COMPTEL distribution available
at that time.  They suggested that COMPTEL, which subtracts background
by imaging with the Compton telescope technique, would be insensitive
to a spatially diffuse component that the large, integrating field of
view of GRIS would accept.  At first glance the \it RHESSI \rm result
seems to support this higher value, since the effective field of view
required to match the GRIS data, 1.04 radians, is identical to the
region I required to be visible for the source intervals.  In future
work, these \it RHESSI \rm source and background fields of view, with
the addition of data from those times in which the central radian is
partially occulted, will be convolved with different model
distributions to provide a normalization for each distribution and to
help choose among them.  Comparison with upcoming results from
INTEGRAL/SPI, which has a smaller field of view, will allow us to test
for the existence of very large-scale diffuse emission.

\acknowledgements

I would like to thank the rest of the \it RHESSI \rm team for
making this work possible, particularly those who worked on
the spectrometer and the data-analysis software.
G. H. Share made valuable suggestions related to this analysis,
and R. P. Lin and H. Hudson thoughtfully reviewed the manuscript.
This work was supported by NASA contract NAS5-98033.

\end{document}